\documentclass[modern]{aastex62}
\received{someday}
\revised{later}
\accepted{not yet}
\submitjournal{ApJ}

\newcommand{\CUBI}{C_{\rm U,B,I}}

\shorttitle{Ruprecht 106}
\shortauthors{Dotter et al.}
\begin{document}

\title{Ruprecht 106: A riddle, wrapped in a mystery, inside an enigma\footnote{Based on observations made with the NASA/ESA Hubble Space Telescope, obtained from the data archive at the Space Telescope Science Institute. STScI is operated by the Association of Universities for Research in Astronomy, Inc. under NASA contract NAS 5-26555.}}

\correspondingauthor{Aaron Dotter}
\email{aaron.dotter@gmail.com}

\author[0000-0002-4442-5700]{Aaron Dotter}
\affiliation{Harvard-Smithsonian Center for Astrophysics}

\author{Antonino P. Milone}
\affiliation{Universit{\`a} di Padova}

\author{Charlie Conroy}
\affiliation{Harvard-Smithsonian Center for Astrophysics}

\author{Anna F. Marino}
\affiliation{Australian National University}

\author{Ata Sarajedini}
\affiliation{Florida Atlantic University}

\begin{abstract}
  Galactic globular clusters (GCs) show overwhelming photometric and spectroscopic evidence for the existence of multiple stellar populations. The question of whether or not there exists a GC that represents a true `simple stellar population' remains open. Here we focus on Ruprecht~106 (R106), a halo GC with [Fe/H]=$-1.5$ and [$\alpha$/Fe]$\simeq0$. A previous spectroscopic study found no sign of the Na-O anticorrelation among 9 of its brightest red giants, which led to the conclusion that R106 is a true simple stellar population GC. Here we present new \emph{Hubble Space Telescope} (HST) Wide Field Camera 3 photometry of R106 that, when combined with archival HST images spanning a 6-year baseline, allows us to create proper motion cleaned color-magnitude diagrams spanning the ultraviolet (F336W) to the near-infrared (F814W). These data allow us to construct the pseudo-color $\CUBI$ that is sensitive to the presence of light-element abundance spreads. We find no evidence of a split along the red giant branch (RGB) in the $\CUBI$ diagram but the width of the RGB ($\sigma_{\CUBI}=0.015$) is marginally broader than expected from artificial star tests ($\sigma_{\CUBI}=0.009$). The observed spread in $\CUBI$ is smaller than any other Galactic GC studied to date. Our results raise important questions about the r{\^o}le of formation environment and primordial chemical composition in the formation of multiple stellar populations in GCs.
\end{abstract}


\section{Introduction} \label{sec:intro}

Observational evidence for the presence of multiple stellar populations (hereafter MPs) in Galactic globular clusters (GCs) is overwhelming \citep{gratton2012,piotto2015,milone2017}. This begs the question: What conditions are required in the formation and early evolution of a GC that allow it form and retain more than one stellar generation to the present day? Insight into this question may be found in searching for GCs that do \emph{not} harbor MPs, if any exist.

In this study we focus on Ruprecht~106 (hereafter R106), which resides in the halo at a Galactocentric radius of $\sim18.5$ kpc \citep{harris1996}. R106 has [Fe/H]=$-1.5$ \citep[][hereafter V13]{brown1997,francois1997,villanova2013} and, remarkably, near solar-scaled abundance ratios among the $\alpha$-capture elements with [O/Fe]$\sim 0$ and [$\alpha$/Fe] $\sim 0$ \citep[][ V13]{brown1997}. The low [$\alpha$/Fe] ratio is unique among Galactic GCs with [Fe/H] $< -1$ \citep{pritzl2005}.

In addition to atypical chemical abundances, R106 has a fairly low mass for a Galactic GC. Mass estimates range from $\log_{10}(M/M_{\odot})=4.77$ \citep{baumgardt2010} to 4.92 \citep{gnedin1997} placing it toward the low end of the Galactic GC mass spectrum.  However, its mass is not anomalously low and \citet{milone2017} have analyzed a handful of GCs with masses lower than R106 that possess strong photometric evidence of MPs. The current mass of any GC is of course only a lower limit on its initial mass, and it is the latter that is likely the most important variable in the formation of MPs \citep[see the discussion of mass budget in][]{renzini2015}.

R106 has previously been found to be 1-3 Gyr younger than other Galactic GCs with similar [Fe/H] \citep{buonanno1990,dacosta1992,buonanno1993,dotter2011}. The age, chemical composition, and location lead to the suggestion that R106 was accreted rather than formed \emph{in situ} \citep{forbes2010}. A recent photometric and spectroscopic study of another likely-accreted GC, IC~4499, found compelling evidence for the presence of MPs \citep{dalessandro2018}. Evidence that intermediate- and old-age GCs in the Small Magellanic Cloud (SMC) also host MPs \citep{dalessandro2016,niederhofer2017,hollyhead2018}. These results indicate that the host galaxy mass, at least down to the level of the SMC, is not a limiting factor in the formation of MPs in GCs.

In the largest spectroscopic study of R106 to date V13 targeted 9 of its brightest red giants. They found [Na/Fe] $< -0.5$: lower than any other Galactic GC but consistent with Local Group dwarf galaxies (see their Figs.~4 and 6). The driving motivation for this paper is that V13 found no evidence of the canonical Na-O anticorrelation in R106. They wrote ``Ruprecht 106 is the first convincing example of a single-population GC (i.e., a true simple stellar population), although the sample is relatively small.'' These results from V13 raise the possibility that the peculiar abundance pattern found in R106 may have some role in the apparent lack of MPs.

Photometry has come to play a major role in the study of MPs, particularly those filters that are sensitive to light-element variations.
The medium-band Str{\"o}mgren pseudo-color $c=(u-v)-(v-b)$ is sensitive to NH variations and, therefore, traces light-element (CNO) variations in GC red giants \citep[see, e.g.,][and references therein]{yong2008,sbordone2011}. \citet{monelli2013} showed the efficacy of constructing a broadband version of $c$, $\CUBI$=(U$-$B)$-$(B$-$I), in the study of MPs. A variety of WFC3 filters have been used to study GCs with aims of untangling the MPs within, including combinations of $F275W$, $F336W$, $F438W$, and $F814W$ to construct the so-called `chromosome map' \citep[][and references therein]{milone2017}. \citet{larsen2014} applied the $F343N$ narrowband filter, which captures the NH feature around 3,300 $\AA$, to the study of GCs. $F343N$ has subsequently been used in combination with broadband filters, see $\S$2 of \citet{bastian2017} for a review of filter combinations used to study MPs.

Inspired by the above we undertook an HST WFC3 observing program to study R106 in F336W (U) and F438W (B) filters, known to be sensitive to light-element abundance variations that are common in GCs \citep{marino2008}. Here we use an HST version of $\CUBI$ with F336W (U), F438W (B), and F814W (I). For convenience we refer to (F336W$-$F438W)$-$(F438W$-$F814W) simply as $\CUBI$. We use these data to test the assertion that R106 is a simple stellar population by searching for broadened sequence---or multiple sequences---in the HST $\CUBI$ color-magnitude diagram (CMD).

\section{Data} \label{sec:data}

\subsection{Observations}

Images used in this paper were obtained through two Hubble Space Telescope observing programs, GO-11586 in Cycle 17 and GO-14726 in Cycle 24.  GO-11856 used the Advanced Camera for Surveys (ACS) Wide Field Channel (WFC) to obtain images of R106 in the F606W and F814W filters spread over 2 sequential orbits; these data were taken in July 2010 and presented by \citet{dotter2011}. GO-14726 used WFC3-UVIS to obtain images of Ruprecht~106 in F336W and F438W filters in 4 non-sequential orbits that overlap with the existing ACS images.  The 4 orbits were spread over several months, between December 2016 and September 2017, in order to obtain images that are rotated by approximately 90-degrees from each other. The rotations were applied in order to best sample the point spread function (PSF) variations across the UVIS chips. Aside from varying the roll angle of the telescope, each orbit included the same pattern of two 1100s exposures in F336W followed by one exposure of 571s in F438W. Visit 1 of the program failed due to problems with guide star acquisition. We applied for and were granted a repeat (HOPR 87495) but, due to scheduling constraints, the decision was made to execute the final two orbits during the same visit and, thus, at the same orientation. Therefore, the 4 successful orbits were taken at 3 distinct orientations instead of 4 as originally planned.

%\begin{deluxetable}{llllll}
%\tablecolumns{6}
%\tablewidth{0pc}
%\tabletypesize{\scriptsize}
%%\rotate
%\tablecaption{GO-11586 Observations \label{obslog}}
%\tablehead{\colhead{Visit}&\colhead{Data set}&\colhead{Date}&\colhead{PA\_V3($^{\circ}$)}&\colhead{F336W}&\colhead{F438W}}
%\startdata
%1\tablenotemark{a} & id8s01*** & 2017-03-29 & 175.3 & $2\times1100$s & 571s \\
%2                  & id8s02*** & 2016-12-10 &  89.7 & $2\times1100$s & 571s \\
%3                  & id8s03*** & 2017-05-30 & 269.7 & $2\times1100$s & 571s \\
%4                  & id8s04*** & 2017-09-21 & 350.0 & $2\times1100$s & 571s \\
%5                  & id8s05*** & 2017-09-21 & 350.0 & $2\times1100$s & 571s
%\enddata
%\tablenotetext{a}{Visit 1 is unusable due to lack of guide star acquisition.  See text for further details.}
%\end{deluxetable}
%
%%
%%  id8s01m5q_flc.fits  id8s01m8q_flc.fits  id8s02b0q_flc.fits  id8s03k6q_flc.fits  id8s03kdq_flc.fits  id8s04ogq_flc.fits  id8s05p8q_flc.fits  id8s05pfq_flc.fits id8s01m6q_flc.fits  id8s02auq_flc.fits  id8s02bpq_flc.fits  id8s03kaq_flc.fits  id8s04oaq_flc.fits  id8s04p2q_flc.fits  id8s05paq_flc.fits

\begin{figure}
\plotone{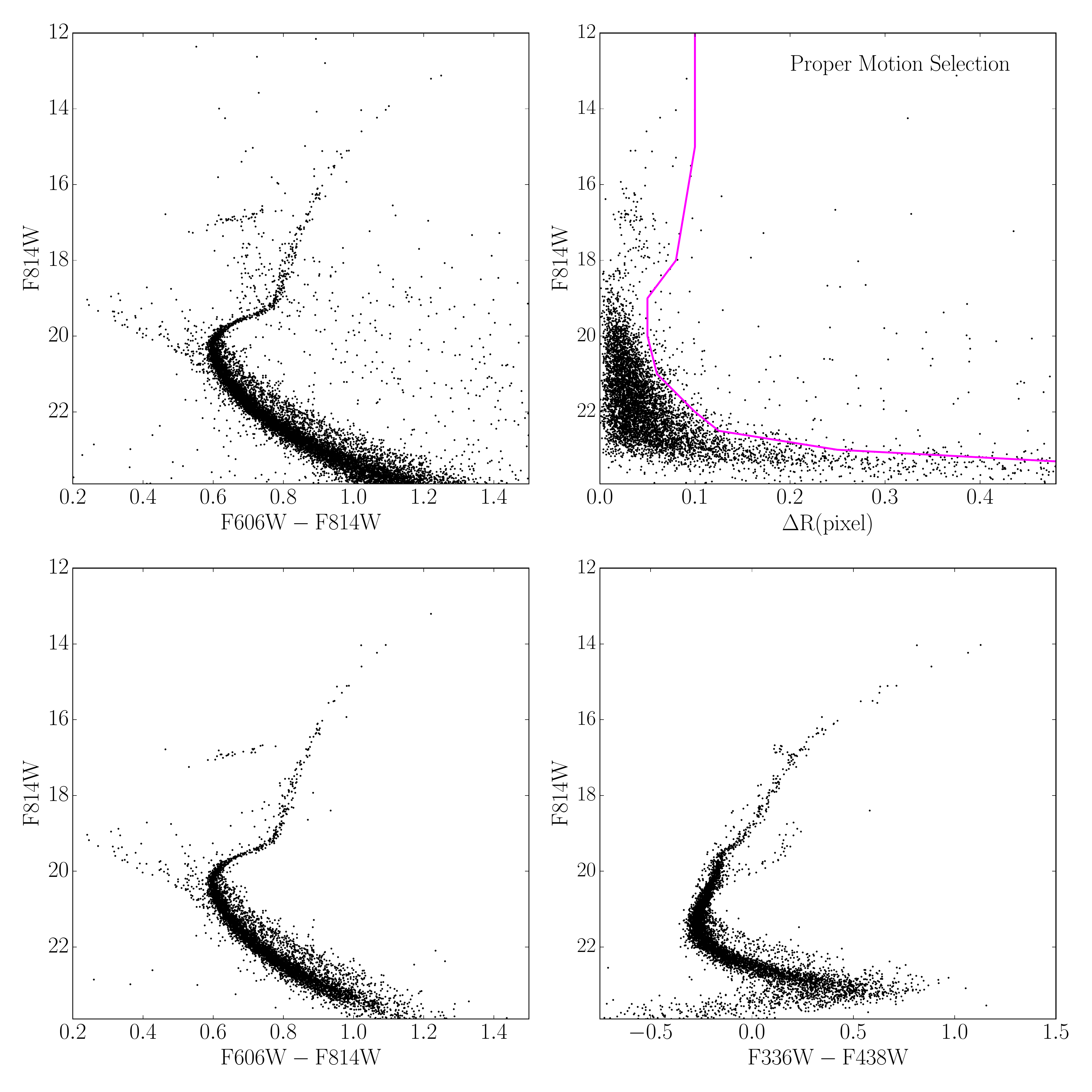}
\caption{(Upper left) The first-epoch F606W$-$F814W CMD of R106 from GO-11586. (Upper right) The proper motion vs. ACS F814W magnitude; the proper motion cut is shown as the solid curve. (Lower left) The ACS F606W$-$F814W CMD showing only proper-motion selected stars. (Lower right) The WFC3 F336W$-$F438W CMD again showing only proper-motion selected stars.\label{fig:PM}}
\end{figure}

\subsection{Photometry and Astrometry}

Astrometry and photometry were measured on flt images using programs developed by \citet{anderson2008b}. The effect of the charge-transfer inefficiency was corrected via the empirical method of \citet{anderson2010}. We derived for each image a $5\times5$ array of perturbation point spread functions (PSFs) using a library of empirical PSFs as well as spatially-varying PSF residuals obtained from isolated stars.

Brighter and fainter stars were measured using different techniques. Brighter stars were measured in each image using the derived PSFs with the program \texttt{img2xym\_wfc3uv}, which is based on \texttt{img2xym\_WFC} \citep{anderson2006} but adapted to WFC3-UVIS. This program also identifies saturated stars and estimates their fluxes and positions. Fainter stars have been measured by using a software package from Jay Anderson that analyzes all the images together to derive the stellar fluxes and luminosities. It is an updated version of the program described by \citet{anderson2008}. Positions of stars were corrected for geometric distortion following \citet{bellini2011}. We also perform artificial star (AS) tests by inserting into the real images and photometering fake stars chosen to lie along the cluster sequence \citep{anderson2008}. These ASs allow us to assess the level of photometric error across the images and over regions of higher and lower stellar density \citep[see $\S$6.2.1 of ][for further details]{milone2009}.

The ACS and WFC3 images are separated by at least 6 years allowing for internal proper motions (PMs) to be determined. Internal PMs are measured within the image frame \{X,Y\} with ${\rm \Delta R = \sqrt{ \Delta X^2 + \Delta Y^2}}$ measured in pixels. In Figure \ref{fig:PM} the upper left panel shows the ACS CMD with no proper motion cut as a solid curve. The upper right panel shows the proper motion, $\Delta$R measured in pixels, plotted against the F814W magnitude along with the proper motion cut as the solid curve. Note that it is more difficult to measure positions of saturated stars due to the difficulty of fitting a PSF to saturated pixels. Stars lying to the left of the proper motion selection curve are shown in F606W$-$F814W in the lower left panel and again in F336W$-$F438W in the lower right panel of Figure \ref{fig:PM}.
The original ACS catalog contains almost 26,800 stars while the proper-motion-selected catalog contains nearly 8,250 stars. The combined ACS and WFC3 catalog---including measurements in all 4 filters---contains just over 6,000 stars.

Since the primary scientific result of this paper relies on the examination of photometric spreads in the CMD, it is reasonable to ask whether or not the observed spread could be due to differential reddening (DR) across the field. The dust extinction maps from \citet{sf11} indicate E(B$-$V)=$0.1499\pm0.0035$ in a $2^{\circ}$ field centered on R106. This translates to E($\CUBI$)$\simeq -0.20\pm0.01$. Applying the differential reddening correction technique described by \citet{milone2012c} we find the maximum spread in the differential correction is $\pm 0.010$ in E(B$-$V) at the 68.27th percentile and full range lowest to highest is 0.087. The average uncertainty in the reddening correction is 0.0036 in E(B$-$V) or 0.005 in E($\CUBI$).

\section{Analysis}

\begin{figure}
  \plotone{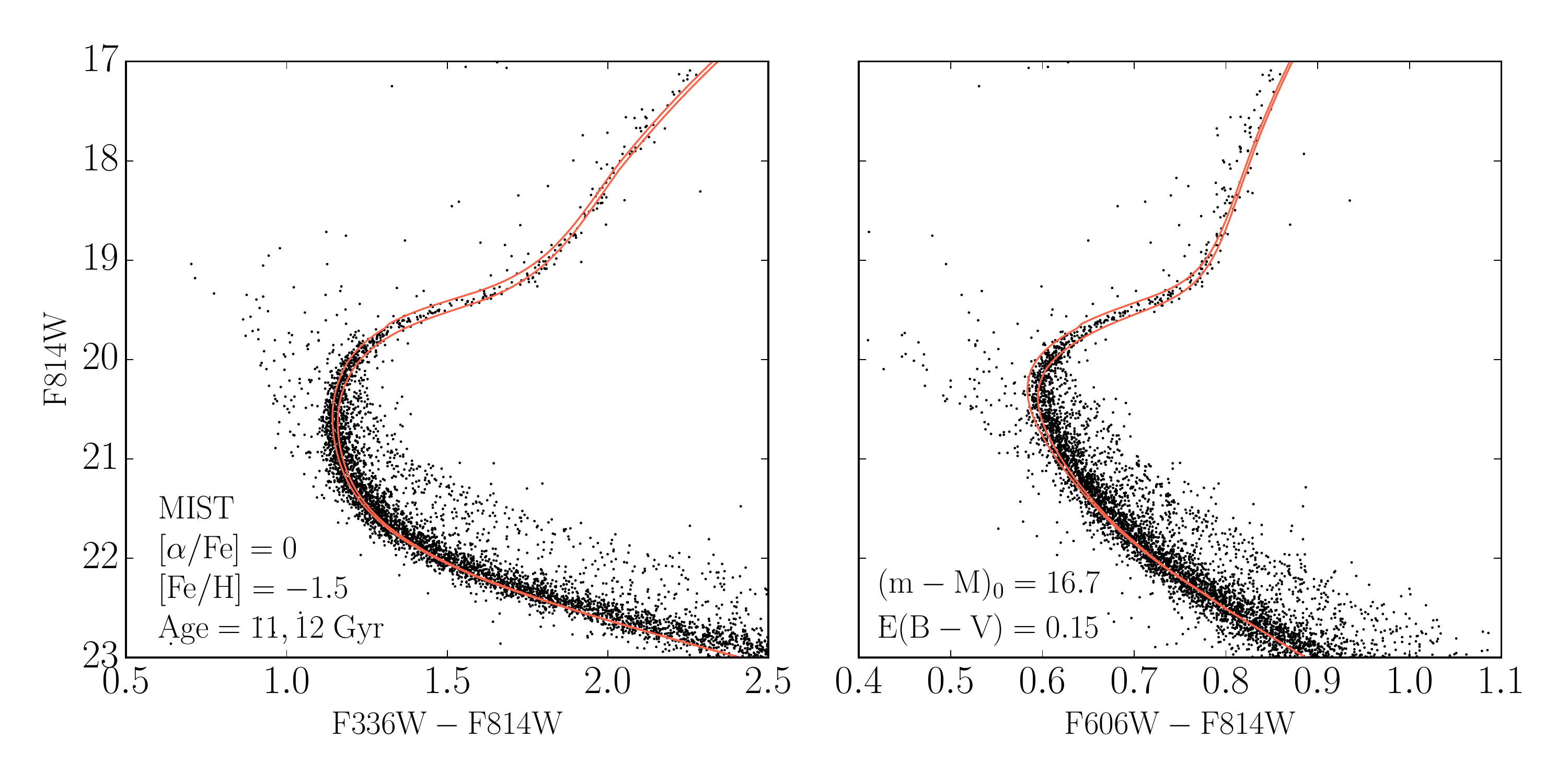}
  \caption{Isochrone comparison with the R106 CMD in F336W$-$F814W (left) and F606W$-$F814W (right). Shown are MIST isochrones \citep{choi2016} with ages of 11 and 12 Gyr, [Fe/H]=$-1.5$, and [$\alpha$/Fe]=0. The MIST models apply an extinction law that reddens each filter self-consistently for a given E(B$-$V).\label{fig:age}}
\end{figure}

One of the characteristics that makes R106 unusual in the Galactic GC population is its relatively young age. \citet{dotter2011} estimated $11.5\pm0.5$ Gyr based on ACS F606W$-$F814W photometry and Dartmouth isochrones \citep{dotter2008}. We confirm this result using both WFC3 and ACS photometry, along with MIST isochrones \citep{choi2016}, in Figure \ref{fig:age}. We have plotted isochrones with ages of 11 and 12 Gyr, with [Fe/H]=$-1.5$ and [$\alpha$/Fe]=0, in both the left and right panels of Figure \ref{fig:age}. The extinction is applied self-consistently for each point along the isochrone using E(B$-$V)=0.15 and the \citet{ccm} extinction curve with ${\rm R_V=3.1}$. Figure \ref{fig:age} shows that the isochrones bracket the stars along the subgiant branch, indicating that the age of R106 lies between 11 and 12 Gyr and we shall use an age of 11.5 Gyr in the following.

\begin{figure}
  \plotone{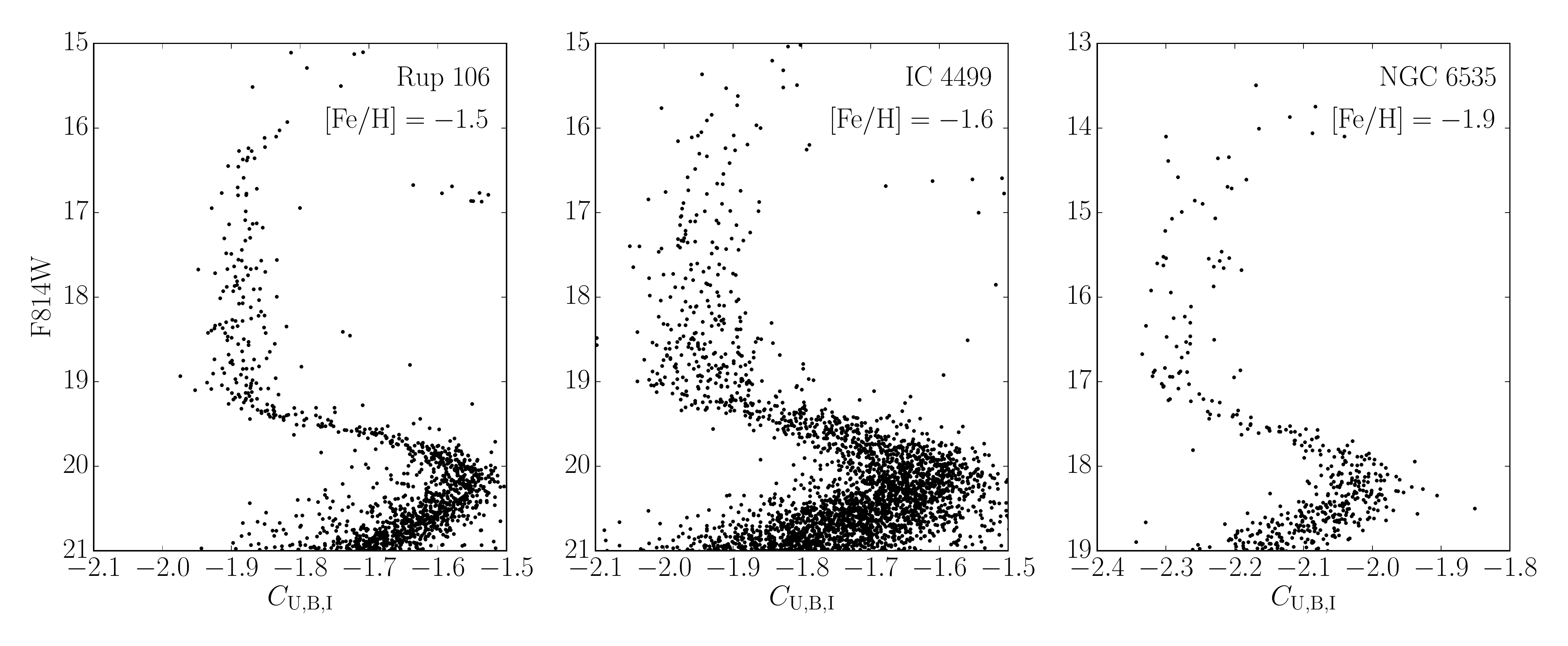}
  \caption{The $\CUBI$ diagrams for R106 (left), IC~4499 \citep[center][]{dalessandro2018} another likely-accreted GC with similar [Fe/H] to R106 and clear evidence of MPs from photometry and spectroscopy, and NGC~6535 \citep[right][]{milone2017} the GC with the smallest RGB spread in the UV Legacy Survey of GCs. All diagrams have been corrected for differential reddening.\label{fig:compare}}
\end{figure}

\begin{figure}
  \plotone{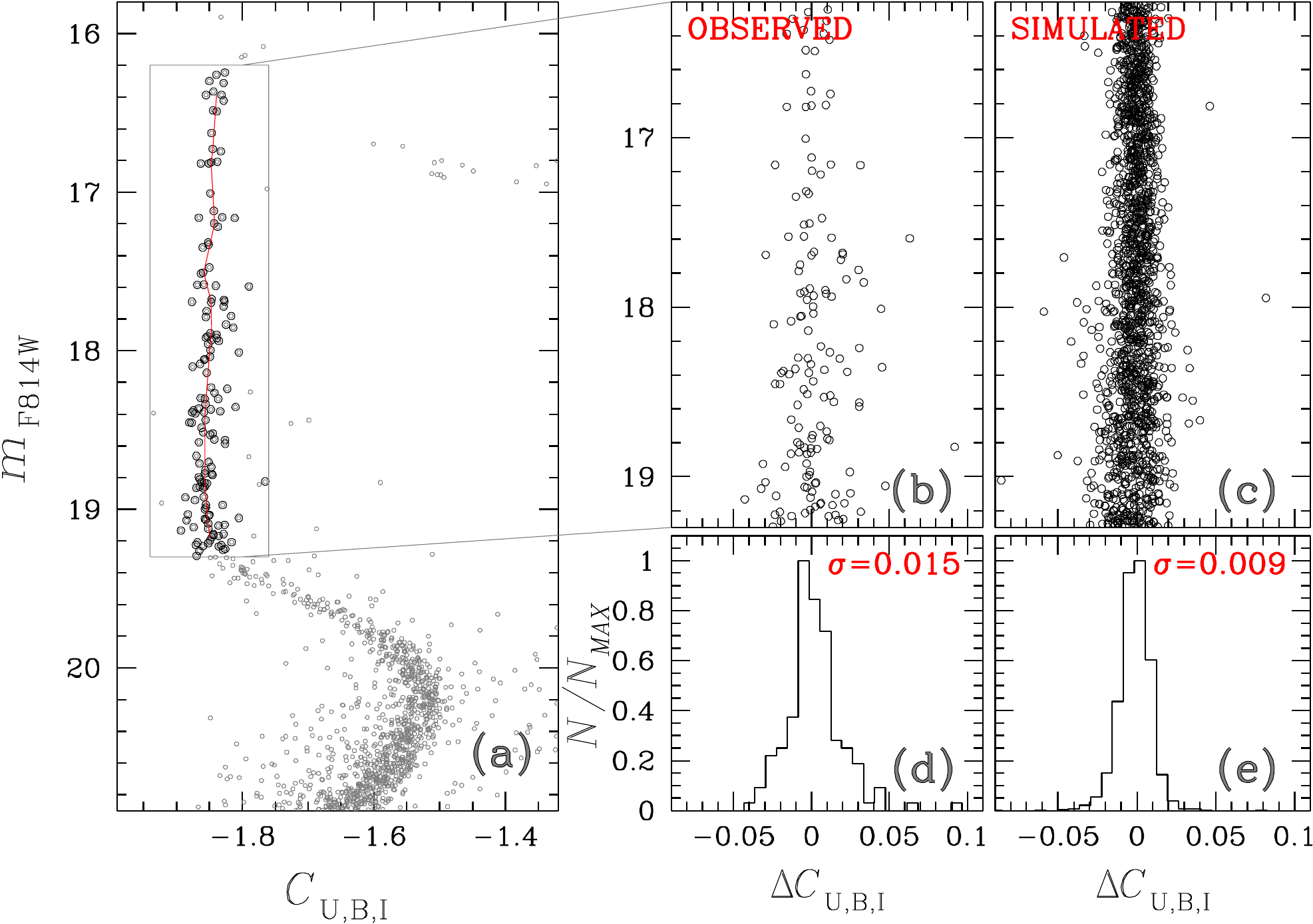}
  \caption{(a) The $\CUBI$ diagram of R106; ridge line is drawn through the RGB is shown in red. (b) Zoom in on the RGB with the color of the ridge line in (a) subtracted from each star. (c) Same as (b) but now showing artificial stars instead of real stars. (d) Histogram of the stars shown in (b). (e) Histogram of the artificial stars shown in (c). See text for more details. \label{fig:AS}}
\end{figure}

As mentioned in $\S$\ref{sec:intro}, the $\CUBI$ diagram provides a useful photometric probe of CNO abundance variations. We begin to explore the R106 $\CUBI$ diagram by comparing it with two other Galactic GCs: IC~4499 \citep{dalessandro2018} and NGC~6535 \citep{milone2017}. IC~4499 has [Fe/H] $\simeq -1.6$ and [$\alpha$/Fe] $\simeq +0.3$ \citep[][and references therein]{dalessandro2018}; it had previously been suggested that IC~4499 did not host MPs but \citet{dalessandro2018} give both photometric and spectroscopic evidence that it does. At [Fe/H] $\simeq -1.92$ and [$\alpha$/Fe] $\simeq+0.4$ \citep{bragaglia2017}, NGC~6535 is more metal-poor than either R106 or IC~4499 and is the GC with the smallest spread in $\CUBI$ among the 57 studied by \citet[][but note that neither R106 nor IC~4499 were included among the 57]{milone2017}. Figure \ref{fig:compare} shows the $\CUBI$ \emph{vs.} F814W diagrams for each of these three GCs. NGC~6535 is both closer to us and more heavily reddened than the other clusters and so both the magnitudes and colors shown are different from the GCs but the ranges are the same. The photometry of all three GCs shown in Figure \ref{fig:compare} has been corrected for differential reddening. The width of the RGB in $\CUBI$ shown in Figure \ref{fig:compare} is smaller for R106 than either of the other two; compare $16 \leq \rm{F814W} \leq 19$ for R106 and IC~4499 or $14 \leq \rm{F814W} \leq 17$ for NGC~6535. To quantify, we have computed the standard deviation of the verticalized $\CUBI$ diagram (see Figure \ref{fig:AS} for an illustration and the accompanying text for details on how this is done) over a 3-magnitude window. The results are: R106 has $\sigma_{\CUBI} = 0.015$ (N=103, see Figure 4) while IC\,4499 has $\sigma_{\CUBI}=0.041$ (N=223) and NGC\,6535 has $\sigma_{\CUBI}=0.033$ (N=49).

In panel (a) of Figure \ref{fig:AS} we show the proper motion cleaned and DR corrected $\CUBI$ diagram. Likely red giants selected in the $F336W-F814W$ vs./ $F814W$ CMD are shown in panel (a) as larger symbols. Panel (b) zooms in on the RGB rectangle highlighted in (a) and ``straightens'' the sequence by subtracting off the color of the ridge line plotted in red in panel (a) to obtain $\Delta \CUBI$. Note that not all stars lying within the rectangle in (a) were selected as likely red giants. Panel (c) shows the same region of the $\CUBI$ diagram but this time plotting the artificial stars that lie in the same region. Panels (d) and (e) show $\Delta \CUBI$ histograms of panels derived from panels (b) and (c) respectively.  Fitting a gaussian to the histograms in panels (d) and (e) results in $\sigma=0.015$ for the R106 stars and $\sigma=0.009$ for the ASs.

In order to quantify the breadth of the RGB in terms of variations in CNO abundances, we used 3 grid of ATLAS \citep{kurucz1970} model atmospheres and SYNTHE \citep{kurucz1993} synthetic spectra to assess the effects of abundance variations on synthetic colors. These grids all have [Fe/H]=$-1.5$ and [$\alpha$/Fe]=0. We then vary C, N, and O, motivated by the range of observed [O/Fe] from V13, while maintaining a constant C+N+O abundance. The first grid has [C/Fe]=0, [N/Fe]=0, and [O/Fe]=0. The second has [C/Fe]=$-0.15$ and [O/Fe]=$-0.15$ while [N/Fe]=$+0.6$ to maintain constant C+N+O. The third grid has [C/Fe]=$-0.30$ and [O/Fe]=$-0.3$ while [N/Fe]=$+0.8$ again to maintain constant C+N+O. The synthetic spectra are then converted into bolometric corrections and applied to the same set of MIST isochrones used in Figure \ref{fig:age}.

We apply these synthetic colors to the same underlying isochrones because CNO variations at constant C+N+O have negligible effect on the ${\rm T_{eff}}$ scale of red giants \citep{dotter2007,vandenberg2012,dotter2015}. However, as shown by \citet{sbordone2011} and \citet{dotter2015}, the effect of CNO variations on spectra blueward of 4,000 $\mathrm{\AA}$, hence also the ultraviolet and blue colors, can be substantial. By comparing isochrones with different mixtures of C, N, and O we can begin to quantify the effect of light-element abundance variations on the R106 $\CUBI$ diagram.  This approach is essentially the same as taken in the study of NGC\,419 by \citet{martocchia2017}.

\begin{figure}
  \plottwo{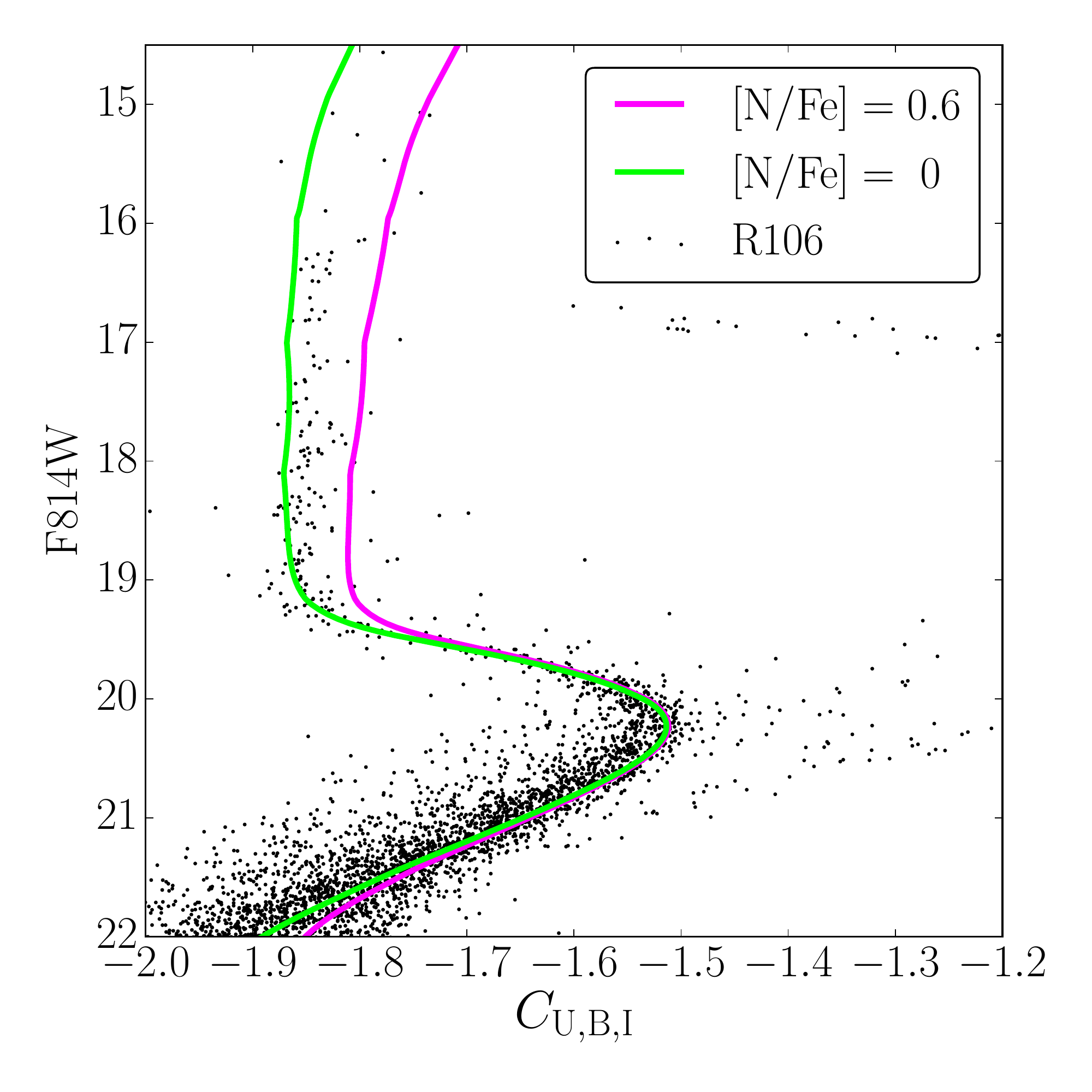}{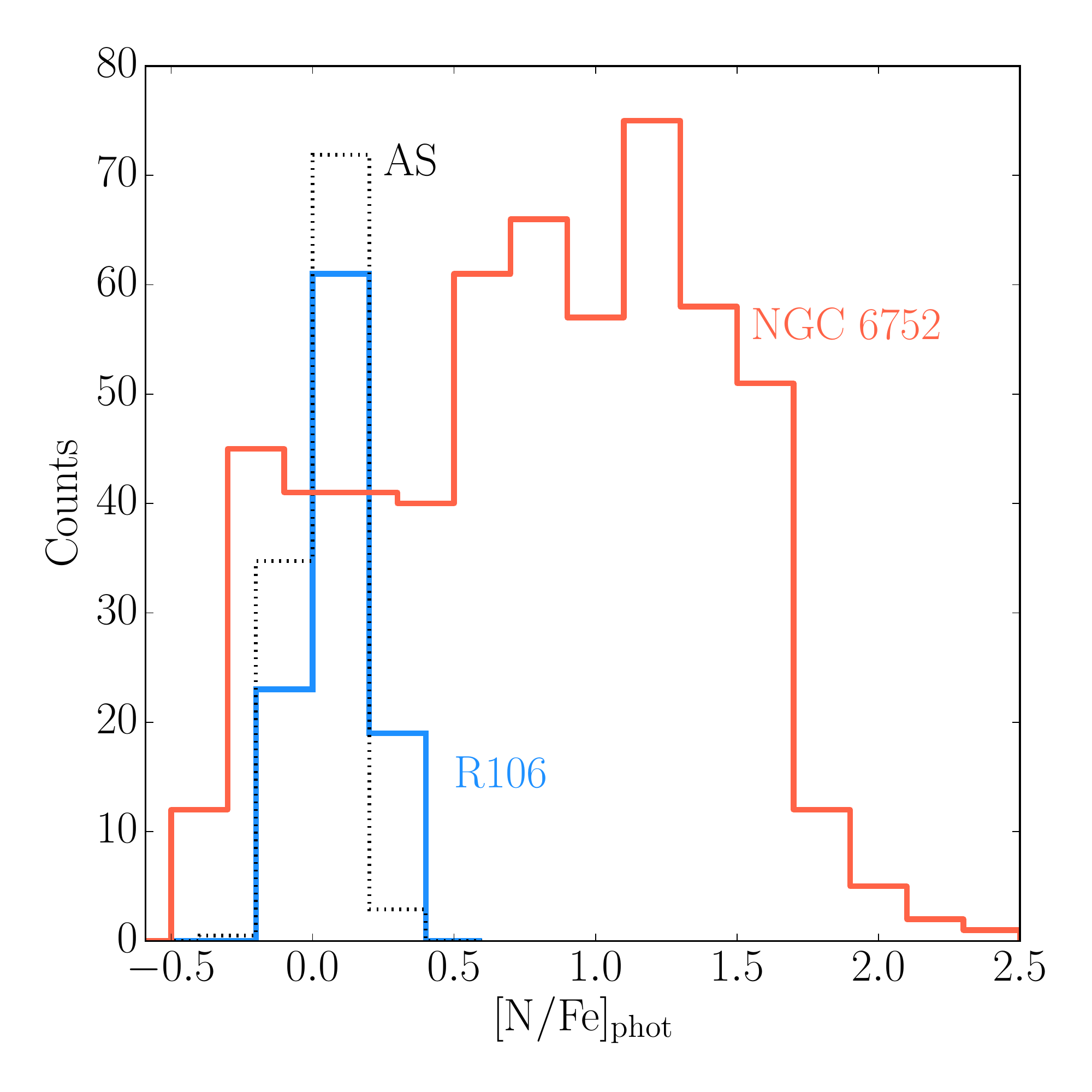}
  \caption{\emph{(Left)} Isochrones with different C, N, and O values plotted against R106 $\CUBI$ data. \emph{(Right)} Using the relationship between $\CUBI$ and [N/Fe] for red giants with $19 \leq $ F814W $\leq 16$ we construct a histogram of photometric [N/Fe] estimates for R106 and compare with similar data for NGC~6752 from \citet{yong2008}. The ASs are shown as the black histogram, normalized to the number R106 RGB stars. \label{fig:models}}
\end{figure}

We compare isochrones at [N/Fe]=0 and +0.6 with the R106 $\CUBI$ diagram in the left panel of Figure \ref{fig:models}. These two isochrones bracket 93 (91\%) of R106 red giants between F814W=19 and F814W=16, resulting in a sample of 103 RGB stars. We use the isochrones at [N/Fe]=0, +0.6, and +0.8 to develop a relation between $\CUBI$, [N/Fe], and F814W over this range of 3 magnitudes and compare with the results obtained for NGC~6752 using Str{\"o}mgren photometry over a luminosity comparable range \citep[][see $\S$4.2 and Fig.\ 8]{yong2008} in the right panel of Figure \ref{fig:models}. A simple, linear version of the relation, which is particular to R106 and assumes the RGB is vertical, is [N/Fe] $\simeq$ $10\times\CUBI + 18.8$ and is valid between $-1.9 \leq \CUBI \leq -1.8$. NGC~6752 makes a compelling comparison because it is an archetypal multiple-population GC that has been studied using spectroscopy and photometry with a strong correlation between Str{\"o}mgren $c$ (or $c_y$) and spectroscopically-determined [N/Fe] \citep{yong2008}. The [N/Fe] distribution of R106 has $\sigma=0.125$. We have also converted the ASs into the [N/Fe] plane using the same procedure and show those as the dotted histogram; the AS distribution has $\sigma=0.085$. While the width of the AS sequence is narrower than that of the real stars, it is important to note that the width of the ASs is only a lower limit and does not account for a number of other factors including, but not limited to, the difference between model and real PSFs, residual effects of DR, and residual field star contamination (even though we have attempted to minimize the last two). In other words, the simple fact that the distribution of RGB stars is wider than the ASs in the $\CUBI$ diagram does not lead us to claim the presence of a real spread in the RGB.

While photometric abundance estimates are neither as accurate nor as precise as spectroscopic ones, the ability to assess the abundance spread among a sample of 103 red giants in R106, and compare it to an even larger sample in another GC, is complementary to the V13 spectroscopic study. Further progress on R106 can be made with HST WFC3 photometry in the F275W filter and spectroscopy targeting C and N abundances in the stars already targeted by V13 as well as increasing the size of the spectroscopic sample.
A final note of caution: Our interpretation of a spread in [N/Fe] relies on the assumption that there \emph{is} some oxygen spread in the cluster, which cannot be excluded based on the V13 spectroscopic study, as well as the assumption that C+N+O is constant. Even if these assumptions are both proven wrong, the observational result of a small spread in $\CUBI$ remains. Further study of R106 is required to determine whether or not the observed spread in $\CUBI$ is due to abundance variations or other factors.

\section{Conclusions} \label{sec:conclusions}
Even before the V13 spectroscopic study R106 was an outlier in the Galactic GC population \citep{pritzl2005}. V13 identified a new dimension in which R106 is peculiar: an apparent absence of the Na-O anti-correlation that has become synonymous with GCs. In this study, we find no compelling evidence for a split RGB and only a marginal spread in $\CUBI$ among RGB stars. Taken together, these findings reinforce the claim by V13 that R106 is the best candidate for a simple stellar population GC.  However, this can only be confirmed through further study of this enigmatic GC.

\acknowledgments
We thank: Robert Kurucz and Fiorella Castelli for maintaining the ATLAS and SYNTHE codes; Emanuele Dalessandro for sharing the IC~4499 photometry catalog; and Ivan Cabrera-Ziri and Diederick Kruijssen for insightful discussions. Support for this work was provided by NASA through grant number HST-GO-14726.004 from the Space Telescope Science Institute, which is operated by AURA, Inc., under NASA contract NAS 5-26555. APM acknowledges support by the European Research Council through the ERC-StG 2016, project 716082 `GALFOR' and by the MIUR through the FARE project R164RM93XW `SEMPLICE'. C.C.\ acknowledges support from the Packard foundation. A.F.M.\ acknowledges support by the Australian Research Council through Discovery Early Career Researcher Award DE160100851.

\vspace{5mm}
\facilities{HST(WFC3), HST(ACS)}

\software{ATLAS,SYNTHE}

%\bibliography{bibliography}

\end{document}